\pdfoutput=1
\documentclass[10pt]{article}

\usepackage{arxiv}
\usepackage{amsmath,amssymb,amsthm}
\usepackage[numbers,sort&compress]{natbib}
\usepackage{booktabs,array,multirow}
\usepackage{graphicx}
\usepackage{xurl}
\usepackage{hyperref}
\usepackage{listings}
\usepackage{xcolor}
\usepackage{enumitem}
\usepackage{float}

\hypersetup{
  colorlinks=true,
  breaklinks=true,
  hypertexnames=false,
  linkcolor=blue!60!black,
  citecolor=blue!60!black,
  urlcolor=blue!60!black
}
\newcommand{\code}[1]{\nolinkurl{#1}}

\newtheorem{lesson}{Lesson}

\definecolor{codebg}{HTML}{F4F5F6}
\runningtitle{Everything Is a Commit: \texttt{git-harness}, a Git-Native Coding Agent Harness}
\headright{Preprint}

\title{Everything Is a Commit:\\\texttt{git-harness}, a Git-Native Coding Agent Harness}

\author{
  Hejian Sang\\[0.3em]
  {\normalfont\small \texttt{https://github.com/HJSang/git-harness}}
}

\date{Technical Report \quad\textbullet\quad November 2026}

\begin{document}
\maketitle

\begin{abstract}
Production coding agent harnesses routinely reinvent five subsets of version control: session trees, file snapshots, context compaction, subagent worktrees, and memory folders.
Because these subsystems live in separate logs and mutable directories, rewinding files desynchronizes conversation history and rewriting prefixes invalidates KV-caches.
We present git-harness, an open-source, zero-dependency coding agent harness that builds directly on Git's object model and plumbing.
Every turn, workspace snapshot, offloaded tool output, and subagent branch becomes an immutable Git object in an isolated namespace, so the host repository stays untouched.
This single store provides atomic conversation-and-code rewind, append-only prompt prefixes, exact recall of compacted history, and single-file Git bundle reproducibility.
On the 89-task Terminal-Bench v2.1 sweep, git-harness matches Pi ($79/89$ vs.\ $76/89$, $p = 0.549$), runs at $68.5\,\mathrm{s}$ vs.\ $72.9\,\mathrm{s}$ median duration, and lowers mean API cost by $16\%$ ($\$0.238$ vs.\ $\$0.282$; median $\$0.150$ vs.\ $\$0.141$).
It adds only $0.55\%$ median Git overhead and exports valid forensic bundles on all 89 runs.
Across 240 cross-harness trials spanning four frontier models, git-harness remains competitive (up to $14/15$ on Opus 5.5 with the updated runtime) with zero Anthropic timeouts and shows where harness designs differ in cost, latency, and failure modes.
Code is available at \url{https://github.com/HJSang/git-harness}.
\end{abstract}

\section{Introduction}

Modern coding agents pair a language model with a runtime harness to run shell commands, edit files, and execute tests~\citep{yang2024sweagent,wang2024openhands,karten2026primeagent}.
Systems such as SWE-agent, OpenHands, Prime Agent, Claude Code, Pi, and Codex already solve realistic containerized engineering tasks~\citep{merrill2026terminalbench}.
Beyond model weights, every harness must manage four kinds of state: the active prompt window, ephemeral tool outputs and worker sandboxes, durable conversation and file history, and cross-session memory.

Source-code audits show that production harnesses differ mainly in prompts and interfaces while repeatedly reinventing the same five subsystems over disconnected stores (Table~\ref{tab:five-subsystems})~\citep{barbaste2026harness}.
This fragmentation causes recurring failures.
Rewinding files without rewinding conversation triggers hallucinated edits, while rewriting earlier prompt prefixes invalidates inference KV-caches on long trajectories.
Dropping failed subagents destroys diagnostic traces, and writing unvalidated notes into mutable folders enables cross-session memory poisoning~\citep{karten2026primeagent}.

Every one of those five subsystems is already a special case of Git's content-addressed object model.
We therefore introduce git-harness, a zero-dependency TypeScript harness built directly on Git's native plumbing (Figure~\ref{fig:architecture}).
Private shadow indices commit conversation events and working trees atomically into isolated Git namespaces without touching \texttt{git status} or user branches.
Because turn commits are immutable and prompt construction walks only first-parent links, second-parent subagent merges preserve worker history while keeping the parent prefix byte-identical for KV-cache reuse.
Oversized tool outputs offload to content-addressed Git blobs, while compacted history stays verbatim searchable through a disposable SQLite index.
Every run also exports a portable Git bundle that survives even hard timeouts.

\begin{table}[!t]
\centering
\caption{\textbf{Five version-control subsystems reinvented by production coding harnesses}~\citep{barbaste2026harness,karten2026primeagent}, contrasted with git-harness's native Git primitives.}
\label{tab:five-subsystems}
\scriptsize
\renewcommand{\arraystretch}{1.15}
\begin{tabular}{@{}p{0.19\linewidth}p{0.42\linewidth}p{0.32\linewidth}@{}}
\toprule
\textbf{Subsystem} & \textbf{Production Harness Implementation} & \textbf{git-harness Primitive} \\
\midrule
Session trees & Custom append-only JSONL trees with parent links (Pi, Claude Code, Codex). & Immutable commit DAG under \code{refs/harness/sessions/*}. \\
Workspace snapshots & Separate pre-edit backup folders (Claude Code \code{file-history/}, \code{backups/}) decoupled from chat logs. & Private \code{GIT\_INDEX\_FILE} shadow tree committed atomically per turn. \\
Context compaction & Lossy LLM summarization rewriting active prompt prefixes (Pi, Claude Code, Codex). & First-parent projection, blob offloading, and disposable FTS5 recall. \\
Subagent isolation & Detached \code{git worktree} checkouts (Claude Code, Codex) or ephemeral workers (Prime Agent) unlinked from parent logs. & Linked worktrees on \code{refs/harness/agents/*} joined via second-parent merge commits. \\
Persistent memory & Mutable, unversioned Markdown folders (Claude Code \code{memory/*.md}, \code{CLAUDE.md}, Codex \code{AGENTS.md}). & Gated commits on \code{refs/harness/memory} and \code{skills} with provenance links. \\
\bottomrule
\end{tabular}
\end{table}

Git Context Controller (GCC)~\citep{wu2025gcc} uses Git only as a metaphor over custom markdown directories inside the workspace, whereas git-harness executes directly on Git's object database and leaves the host repository clean.
Prime Agent~\citep{karten2026primeagent} pairs a persistent Python REPL with custom disk persistence and unvalidated online skill updates.
In contrast, git-harness enforces atomic conversation-and-file commits, deterministic pre-commit gates, and append-only prompt prefixes at the storage layer.
Every run exports per-step timing and token telemetry alongside its Git bundle, which helps separate harness bugs from model capability limits~\citep{sghaier2026dontblame,zhang2026disclose}.

We evaluate git-harness to test whether a Git-native storage layer is practical and competitive alongside established harness designs.
On the 89-task Terminal-Bench v2.1 sweep~\citep{merrill2026terminalbench}, git-harness matches Pi ($79/89$ vs.\ $76/89$, $p = 0.549$) with $68.5\,\mathrm{s}$ vs.\ $72.9\,\mathrm{s}$ median duration and $16\%$ lower mean API cost.
Git checkpointing adds only $0.55\%$ median wall time, and all 89 runs export valid bundles with zero harness crashes.
A four-arm ablation on 15 feature-active tasks isolates blob offloading, stuck-loop guardrails, and Turn-0 subagents; disabling always-on subagents cuts median cost by $55\%$.
Across 240 matrix trials spanning four frontier models and five harnesses, git-harness achieves comparable Pass@1 with zero Anthropic timeouts and shows how harness choices shape cost, latency, and failure modes (Table~\ref{tab:matrix}).

\section{Architecture}
\label{sec:architecture}

\begin{figure}[!t]
\centering
\includegraphics[width=\linewidth]{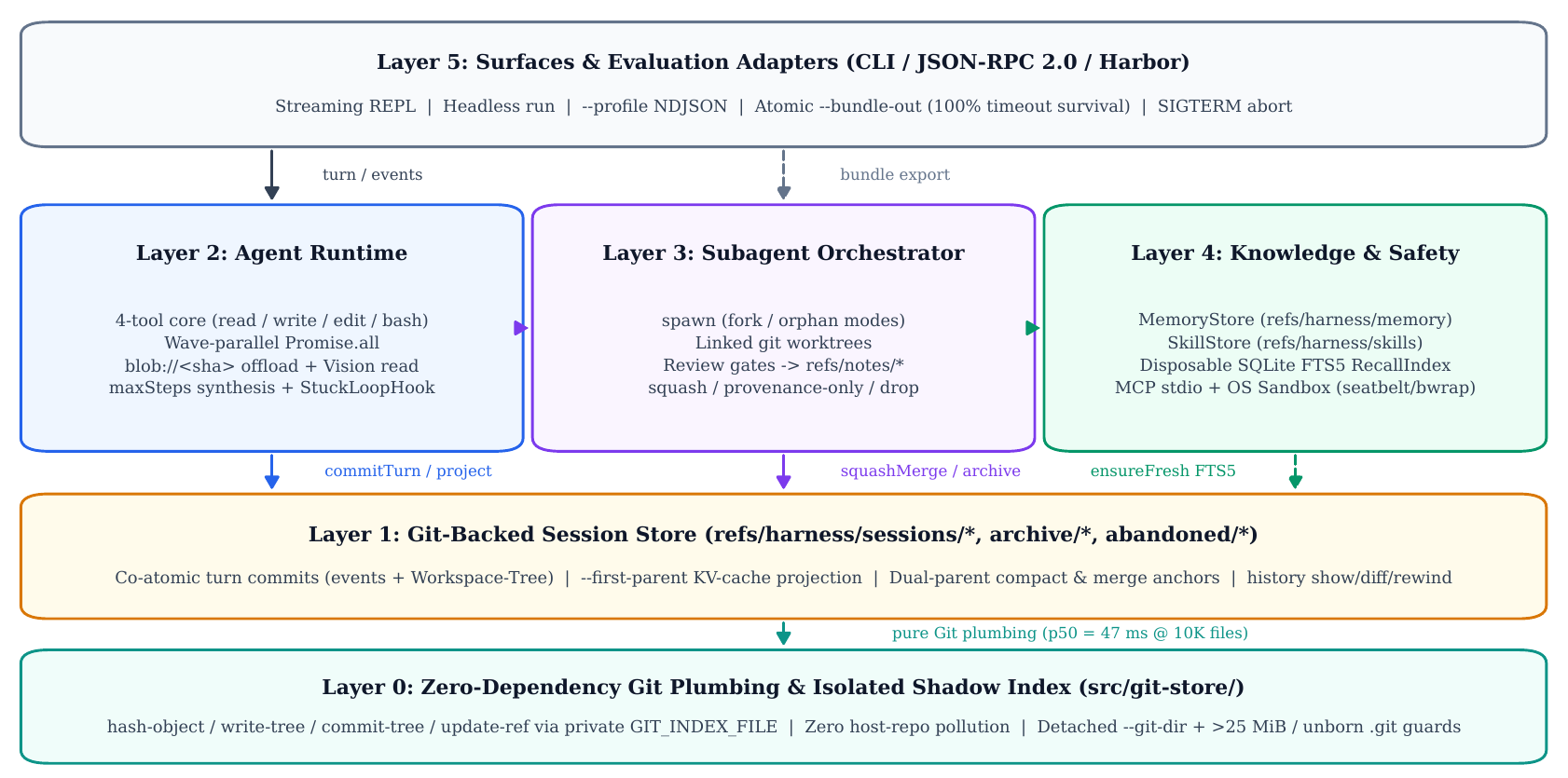}
\caption{\textbf{Six-layer acyclic architecture of git-harness.} Layers~0--1 ground every session turn and workspace snapshot in Git plumbing, while Layers~3--4 attach to the minimal four-tool Layer~2 runtime exclusively through narrow tool and hook interfaces, keeping initial prompt overhead compact without polluting the host repository.}
\label{fig:architecture}
\end{figure}

Figure~\ref{fig:architecture} shows the six-layer architecture of git-harness.
Lower layers own Git storage and session commit graphs (\S\ref{sec:arch-overview}--\ref{sec:git-core}); middle layers execute tools, offload blobs, and coordinate worktree subagents (\S\ref{sec:runtime-orch}); upper layers handle cold search, memory, and operational guardrails (\S\ref{sec:knowledge}--\ref{sec:invariants}).

\subsection{Six-layer acyclic architecture}
\label{sec:arch-overview}
git-harness is organized into six strictly acyclic layers where lower layers never import higher ones.
Layer~0 handles Git object storage, Layer~1 manages session DAGs and prompt projection, and Layer~2 runs the core tool loop and hooks.
Layer~3 coordinates worktree subagents, Layer~4 provides cold-recall search and versioned memory, and Layer~5 exposes CLI and RPC surfaces.
Layers~3 and~4 attach to Layer~2 only through narrow tool and hook interfaces.
This separation keeps the default prompt compact when optional features are disabled.
The runtime has zero npm dependencies and runs natively on Node.js ($\ge 22.18$).
It uses built-in SQLite for disposable search indexing and system \texttt{git} for all persistent storage.

\subsection{Mapping agent state onto Git's object model}
\label{sec:state-hierarchy}
\begin{figure}[!t]
\centering
\includegraphics[width=\linewidth]{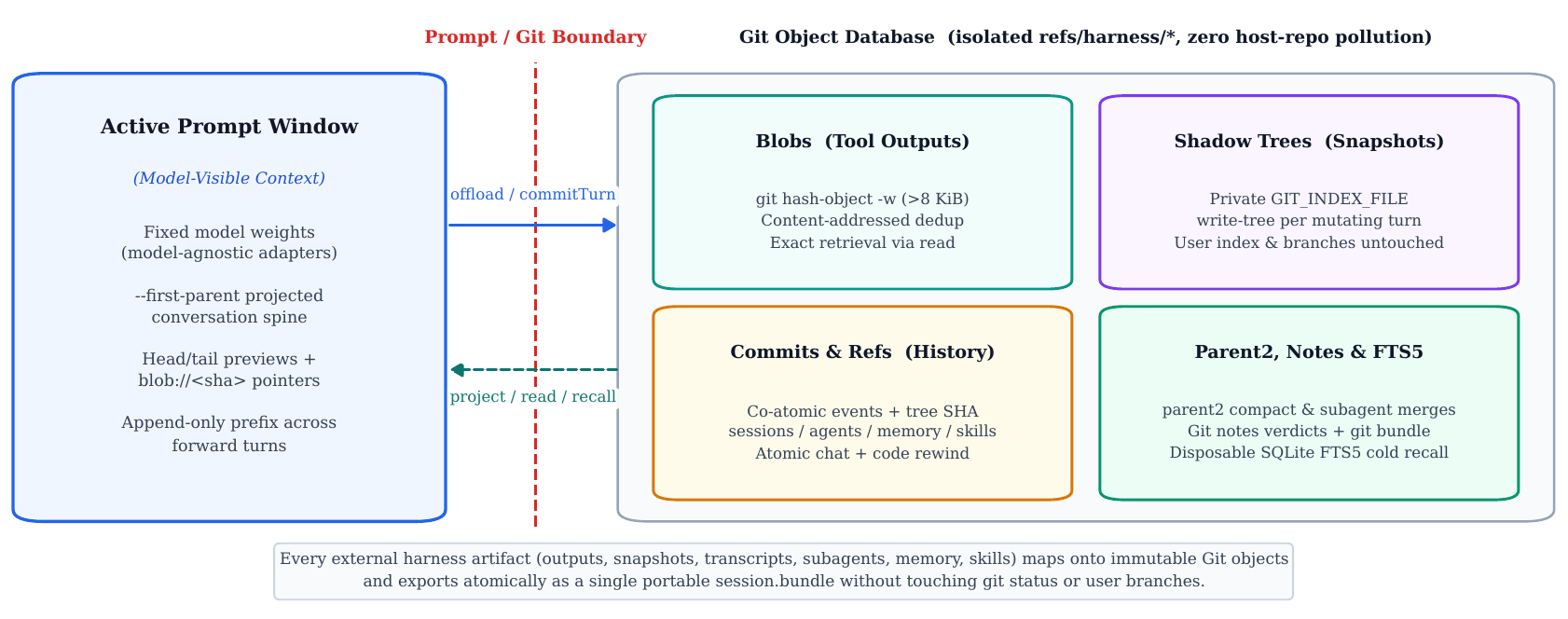}
\caption{\textbf{Git-backed agent state organization.} Explicit transitions move data between the active prompt window and durable Git objects (blobs, shadow trees, commits, refs, and notes) without mutating prior prompt prefixes or polluting the host repository.}
\label{fig:state-hierarchy}
\end{figure}

A single LLM call conditions only on fixed weights and its active prompt window.
Everything else lives outside the model.
git-harness maps each external state tier directly onto a native Git primitive (Figure~\ref{fig:state-hierarchy}):
\begin{itemize}[leftmargin=*,nosep]
  \item \textbf{Blobs (oversized tool outputs):} Large outputs go straight to Git's object database. The active prompt receives only a short preview and a blob pointer.
  \item \textbf{Trees (private workspace snapshots):} Each mutating turn captures the working directory through a private shadow index. The user's staging area and branches remain untouched.
  \item \textbf{Commits and Refs (sessions, subagents, memory, skills):} Conversation events and workspace trees share one immutable commit on isolated refs. Model-agnostic trailers allow swapping adapters mid-session.
  \item \textbf{Second parents, Notes, and Bundles (provenance and portability):} Dual-parent merges link compacted or delegated history without polluting the first-parent spine. Git notes store verification verdicts, and a disposable SQLite FTS5 index rebuilds cold search directly from Git objects.
\end{itemize}

\subsection{Shadow-index plumbing, atomic DAG, and append-only prefixes}
\label{sec:git-core}
\paragraph{Private shadow index (Layer 0).}
Standard Git commands mutate \texttt{.git/index} and \texttt{HEAD}, colliding with user staging areas and branches.
Layer~0 avoids this by passing a private index file via \texttt{GIT\_INDEX\_FILE}.
It runs \texttt{git add -A} and \texttt{git write-tree} into that shadow index without touching user branches.
For non-Git workspaces such as containerized benchmarks, detached \texttt{--git-dir} mode provides full versioning with zero \texttt{.git} footprint.
Layer~0 also guards against nested unborn \texttt{.git} directories and excludes compiled artifacts, VM disk images, and files over $25\,\mathrm{MiB}$.
On read-only steps, it skips workspace hashing and reuses the parent commit's tree SHA in $3\text{--}8\,\mathrm{ms}$.

\paragraph{Atomic session commits (Layer 1).}
Each turn commit stores the turn's events, state metadata, the exact post-step working tree, and an optional summary on compaction or subagent merges.
Structured Git trailers record the model name, token counts, cache hits, and workspace tree SHA.
Because conversation and files share one immutable commit, rewinding to any commit SHA restores both atomically while keeping the discarded path on an abandoned branch.

\paragraph{Append-only prompt prefixes.}
Let $H_t$ be the tip of the session branch after step $t$.
Prompt projection $\Pi(H_t)$ walks \texttt{git rev-list --first-parent} back to the nearest compaction anchor or root.
Committed Git objects are immutable, forward turns append along the first parent, and subagent merges attach via the second parent.
As a result, $\Pi(H_t)$ stays a byte-identical prefix of $\Pi(H_{t+1})$ across every forward non-compacting step, given a fixed system prompt, tool schema, and offload previews.
Actual cache hits still depend on provider-side TTL and eviction, which lie outside the harness.

\subsection{Wave-parallel runtime, blob offloading, and subagent orchestration}
\label{sec:runtime-orch}
\paragraph{Minimal four-tool runtime and optional stateful Python REPL (Layer 2).}
Layer~2 exposes four core tools (\texttt{read}, \texttt{write}, \texttt{edit}, \texttt{bash}) with wave-parallel execution.
Contiguous read-only or parallel-safe calls run concurrently with deterministic output ordering, while mutating calls serialize the wave.
Adopting the persistent Python REPL idea from Prime Agent~\citep{karten2026primeagent}, an optional \texttt{python\_repl} tool keeps heavy libraries (PyTorch, NumPy, SciPy, Stan) resident across turns and restarts automatically on timeout or crash.
Two lightweight guards protect long runs: a per-step streaming watchdog aborts hung model responses, and a pre-exit Git diff hook catches stray temporary files.
The \texttt{read} tool handles line ranges, multimodal images, historical commit files, skill documents, and offloaded blobs.
Any tool output over a threshold ($8\,\mathrm{KiB}$ default for \texttt{bash}) is written to Git's object store and replaced in the prompt with a head-and-tail preview plus a blob pointer.
This combines observation masking~\citep{lindenbauer2025complexity} with exact on-demand retrieval.

\paragraph{Git-native subagent lifecycle (Layer 3).}
Subagent orchestration follows four stages:
\begin{enumerate}[leftmargin=*,nosep]
  \item \textbf{Spawn:} Creates an isolated subagent branch, either forked from parent context or started as an orphan. It can also provision a linked worktree seeded with the parent's uncommitted files.
  \item \textbf{Run:} Executes either the native runtime loop or an external CLI runner under one interface.
  \item \textbf{Review:} Runs shell verification gates inside the subagent's worktree and stores structured verdicts in Git review notes.
  \item \textbf{Integrate:} Acquires a per-session lock to prevent ref races across parallel workers. It supports squash-merging with three-way conflict checks and dual-parent provenance, cherry-picking, or archiving dropped branches.
\end{enumerate}

\paragraph{Boundary-only agent-to-agent communication.}
Some harnesses coordinate running workers through live message queues or shared chat buses~\citep{karten2026primeagent}.
git-harness instead mediates agent-to-agent communication strictly at Git branch boundaries (spawn and integrate).
Parent context and uncommitted files flow downward at spawn time.
Verification verdicts and compact summaries flow upward through Git notes and dual-parent merge commits, while dropped worker branches remain searchable via cold recall.
This design gives up mid-flight peer steering.
In return, it never rewrites the parent prompt prefix, serializes merges without ref races, and preserves complete commit-level auditability.

\subsection{Exact retrieval of compacted history and gated memory/skills}
\label{sec:knowledge}
Persistent project memory and versioned skills reside on dedicated orphan Git branches.
Every proposed update passes deterministic pre-commit gates that reject leaked API keys, private keys, malformed frontmatter, or oversized files.
Each accepted commit records a second-parent link back to its originating session commit, which directly addresses cross-session memory contamination~\citep{karten2026primeagent}.
Exact retrieval of compacted history (\emph{cold recall}) uses a disposable SQLite FTS5 index over all harness branches.
This index can be wiped and deterministically rebuilt from Git objects at any time.

\subsection{Operational lessons for long-horizon runs}
\label{sec:invariants}
Running git-harness across hundreds of long-horizon tasks revealed four practical engineering lessons enforced across Layers~1--5:

\begin{lesson}[Graceful Synthesis Handoff on Step Budget]
Action step budgets should be decoupled from final handoff synthesis.
When an agent hits its step limit mid-tool-loop, hard-stopping discards verified findings in favor of pre-tool narration.
git-harness instead runs one final tool-disabled synthesis call, commits it to the branch, and reports the step-limit stop reason explicitly.
\end{lesson}

\begin{lesson}[Complete Cold-Storage Indexing]
Archiving or rewinding removes a branch from active first-parent projection, never from cold search.
The FTS5 index covers active sessions, subagents, archived and abandoned branches, and memory.
Before each recall query, it checks a fingerprint of current harness ref tips and rebuilds only when a ref has changed.
\end{lesson}

\begin{lesson}[Monotonic Oscillation Guardrails and Timeout Forensics]
A sliding-window churn detector that fires on multiples of $k$ silently stalls once edits fill window $W$ with $W \bmod k \neq 0$.
For example, $W = 10$ and $k = 3$ saturate at count $10$ and never hit $12$.
Our stuck-loop hook pairs sliding-window density with a monotonic per-path edit counter plus step-budget and wall-clock notices.
It also gates subagent recovery hints on whether subagents are enabled.
To preserve forensics on hard timeouts, git-harness atomically refreshes its Git bundle on commit~1, every 10 commits, compaction, \texttt{SIGTERM}, and exit.
\end{lesson}

\begin{lesson}[Wire-Protocol Fidelity and Prefix-Invariant Image Bounding]
Interleaved reasoning models need prior reasoning content echoed on assistant messages; otherwise they re-derive state on every turn.
Adaptive-thinking models require high streaming output ceilings ($64\mathrm{K}$ auto-SSE on Anthropic) and strict argument validation on \texttt{write} so truncated calls never coerce to empty files.
They also need continuation recovery after hitting output limits and explicit refusal telemetry.
Finally, capping inline multimodal images at $4$ per compaction window prevents decode collapse without mutating earlier prompt prefixes.
\end{lesson}

\section{Evaluation}
\label{sec:evaluation}

Our evaluation addresses four research questions:
\begin{itemize}[leftmargin=*,nosep]
  \item \textbf{RQ1 (Overhead and prefix reuse):} Does native Git plumbing scale to large repositories at low overhead while preserving append-only prompt prefixes on long-horizon tasks?
  \item \textbf{RQ2 (Single-seed component ablation):} How do blob offloading, stuck-loop guardrails, and always-on subagent delegation affect latency, cost, and timeouts?
  \item \textbf{RQ3 (Cross-harness comparison):} How do five harness architectures (git-harness, Pi, Prime Agent, Claude Code, Codex) differ in accuracy, Anthropic prompt-cache economics, tool usage, and failure modes across four frontier models?
  \item \textbf{RQ4 (Debugging lessons and unpacked DAGs):} What failure modes did Git bundle forensics uncover during development, and what do unpacked commit graphs look like in practice?
\end{itemize}

Following recent evaluation standards~\citep{kapoor2026hal,zhu2025abc}, we report paired non-parametric tests (exact McNemar and Wilcoxon signed-rank), bootstrap confidence intervals, and raw-log-verified API costs.

\subsection{RQ1: Scaling microbenchmarks, synthetic ablations, and full 89-task sweep}
\label{sec:rq1}

\begin{figure}[!t]
\centering
\includegraphics[width=0.62\linewidth]{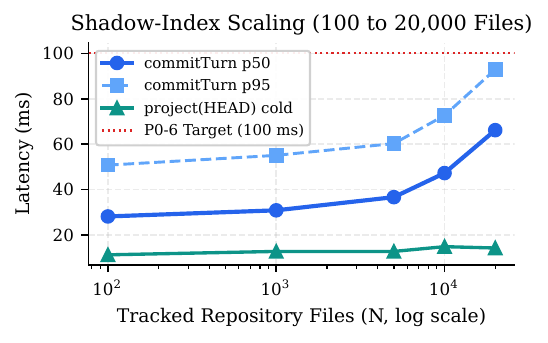}
\caption{\textbf{Shadow-index scaling.} Persistent \texttt{GIT\_INDEX\_FILE} \texttt{commitTurn} latency remains below $100\,\mathrm{ms}$ ($p_{95}$) at $20{,}000$ tracked files while \texttt{project(HEAD)} remains flat.}
\label{fig:scaling}
\end{figure}

\paragraph{Repository scaling.}
Shadow-index turn commits stay under $100\,\mathrm{ms}$ ($p_{95}$) even at $20{,}000$ tracked files plus $500$ ignored files (Table~\ref{tab:latency}, Figure~\ref{fig:scaling}).
With workspace mutations on every turn, median commit latency scales from $28.2\,\mathrm{ms}$ at $100$ files to $66.2\,\mathrm{ms}$ at $20{,}000$ files.
Warm prompt projection stays flat at $0.1\,\mathrm{ms}$ regardless of repository size.

\begin{table}[!t]
\centering
\caption{Git shadow-index latency scaling ($100$ to $20{,}000$ tracked files + $500$ ignored files).}
\label{tab:latency}
\small
\begin{tabular}{@{}rrrrr@{}}
\toprule
\textbf{Tracked Files ($N$)} & \textbf{\texttt{commitTurn} p50} & \textbf{\texttt{commitTurn} p95} & \textbf{\texttt{project} Warm} & \textbf{\texttt{project} Cold (21 msgs)} \\
\midrule
$100$      & $28.2\,\mathrm{ms}$ & $50.8\,\mathrm{ms}$ & $0.1\,\mathrm{ms}$ & $11.3\,\mathrm{ms}$ \\
$1{,}000$  & $30.9\,\mathrm{ms}$ & $55.1\,\mathrm{ms}$ & $0.1\,\mathrm{ms}$ & $12.8\,\mathrm{ms}$ \\
$5{,}000$  & $36.7\,\mathrm{ms}$ & $60.3\,\mathrm{ms}$ & $0.1\,\mathrm{ms}$ & $12.8\,\mathrm{ms}$ \\
$10{,}000$ & $47.3\,\mathrm{ms}$ & $72.7\,\mathrm{ms}$ & $0.1\,\mathrm{ms}$ & $14.9\,\mathrm{ms}$ \\
$20{,}000$ & $66.2\,\mathrm{ms}$ & $92.9\,\mathrm{ms}$ & $0.1\,\mathrm{ms}$ & $14.3\,\mathrm{ms}$ \\
\bottomrule
\end{tabular}
\end{table}

\paragraph{Constructed synthetic P1 ablations (\texttt{npm run eval}).}
Four deterministic synthetic rubric scenarios isolate each architectural mechanism from LLM decoding variance (Table~\ref{tab:p1}).
Active prompt size drops by $76.5\%$ compared to unmasked linear transcripts.
Multi-compaction KV-cache retention rises from $47.5\%$ (an $8$-message sliding window) to $91.0\%$.
Cold recall after three compactions rises from $10\%$ (active context only) to $100\%$, while subagent handoffs compress context to $3.0\%$ and preserve full \texttt{parent2} provenance.

\begin{table}[!t]
\centering
\caption{Synthetic P1 ablation scorecard (\texttt{npm run eval}).}
\label{tab:p1}
\small
\begin{tabular}{@{}llcc@{}}
\toprule
\textbf{ID} & \textbf{Metric} & \textbf{\texttt{git-harness}} & \textbf{Baseline} \\
\midrule
\texttt{P1-POLLUTION} & Active Context Reduction & \textbf{$-76.5\%$} ($1{,}054$ ch) & $0.0\%$ ($4{,}489$ ch) \\
\texttt{P1-KVCACHE}   & KV-Cache Prefix Hit (25 turns + compact) & \textbf{$91.0\%$} & $47.5\%$ (sliding window) \\
\texttt{P1-RECALL}    & Cold Recall Precision (after 3 compactions) & \textbf{$100\%$} ($10/10$) & $10\%$ ($1/10$ active-only) \\
\texttt{P1-HANDOFF}   & Subagent Handoff Compression Ratio & \textbf{$3.0\%$} + \texttt{parent2} & $100\%$ (inline transcript) \\
\bottomrule
\end{tabular}
\end{table}

\begin{figure}[!t]
\centering
\includegraphics[width=\linewidth]{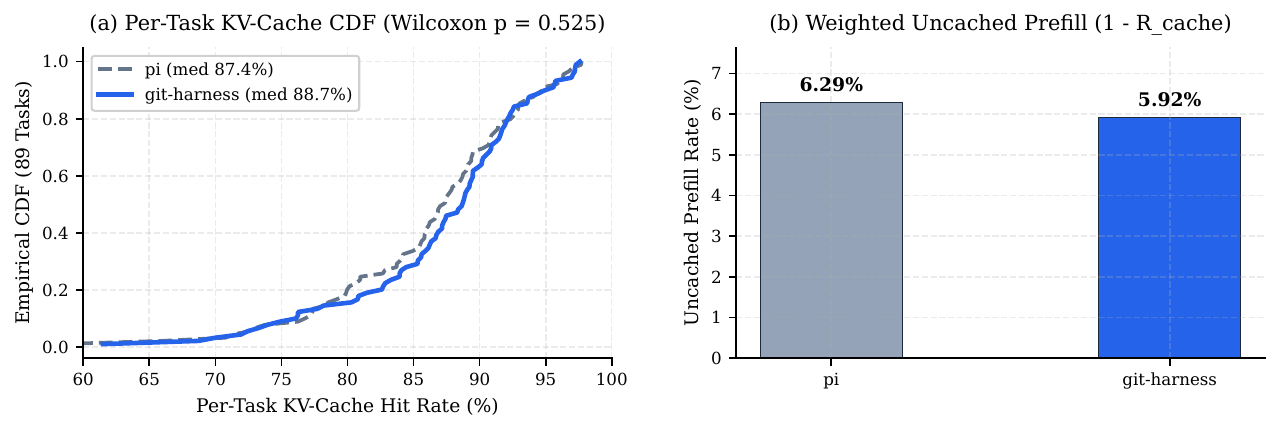}
\caption{\textbf{KV-cache efficiency across 89 Terminal-Bench v2.1 tasks.} (a) Empirical CDF of per-task KV-cache hit rate (median $88.72\%$ vs.\ $87.39\%$ for Pi, paired Wilcoxon $p = 0.525$); (b) weighted uncached prefill rate ($1 - R_{\text{cache}}$), $5.92\%$ for git-harness vs.\ $6.29\%$ for Pi.}
\label{fig:kvcache}
\end{figure}

\begin{figure}[!t]
\centering
\includegraphics[width=\linewidth]{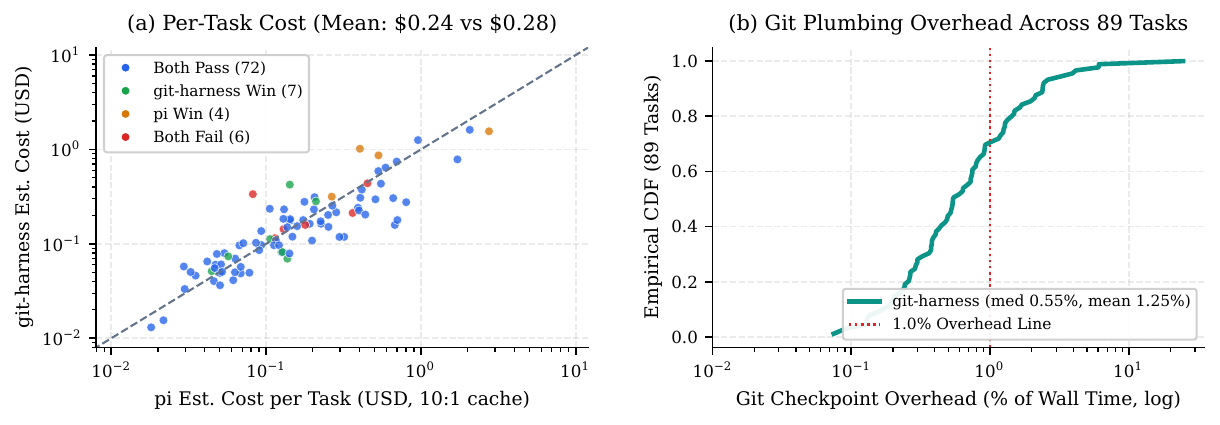}
\caption{\textbf{Cost and Git overhead across 89 tasks.} (a) Paired estimated API cost under 10:1 cache pricing (mean $\$0.238$ vs.\ $\$0.282$, total $\$21.16$ vs.\ $\$25.14$); (b) empirical CDF of Git checkpoint overhead (\% wall time), with median $0.55\%$ and mean $1.25\%$.}
\label{fig:cost}
\end{figure}

\paragraph{Full 89-task Terminal-Bench v2.1 sweep.}
We evaluated git-harness against Pi on all 89 Terminal-Bench v2.1 tasks~\citep{merrill2026terminalbench} ($k=1$) using an identical OpenAI-compatible reasoning endpoint, identical timeouts, and identical Harbor sandboxes (Table~\ref{tab:tb21main}, Figures~\ref{fig:kvcache} and~\ref{fig:cost}).

\begin{table}[!t]
\centering
\caption{Clean full 89-task Terminal-Bench v2.1 sweep ($k=1$, identical reasoning model and timeouts; 10:1 cache pricing: $\$3$/1M uncached input, $\$0.30$/1M cached input, $\$15$/1M output).}
\label{tab:tb21main}
\small
\begin{tabular}{@{}lcc@{}}
\toprule
\textbf{Metric} & \textbf{\texttt{pi}} & \textbf{\texttt{git-harness}} \\
\midrule
Tasks Passed (Clean Full-Sweep Pass@1) & $76 / 89$ ($85.39\%$) & \textbf{$79 / 89$ ($88.76\%$)} \\
McNemar Exact Two-Sided $p$ & --- & $p = 0.549$ ($7$ vs.\ $4$) \\
Failures (Logic / AgentTO / VerifTO / Crash) & $10$ / $2$ / $1$ / $0$ & \textbf{$9$ / $1$ / $0$ / $0$} \\
Agent Duration (Median / Mean) & $72.9\,\mathrm{s}$ / $231.2\,\mathrm{s}$ & \textbf{$68.5\,\mathrm{s}$ / $194.6\,\mathrm{s}$} \\
Est.\ API Cost (Median / Mean, 10:1 cache) & $\$0.141$ / $\$0.282$ & $\$0.150$ / \textbf{$\$0.238$} \\
Per-Task KV-Cache Hit (Mean / Median) & $86.19\%$ / $87.39\%$ & \textbf{$86.95\%$ / $88.72\%$} \\
Weighted Live KV-Cache Hit Rate & $93.71\%$ & \textbf{$94.08\%$} \\
Git Checkpoint Overhead (Median / Mean) & --- & \textbf{$0.545\%$ / $1.248\%$} \\
Portable \texttt{session.bundle} Survival & --- & \textbf{$89/89$ ($100\%$, $42.5\,\mathrm{KB}$ med)} \\
\bottomrule
\end{tabular}
\end{table}

git-harness solves $79/89$ ($88.8\%$) tasks compared to Pi's $76/89$ ($85.4\%$) with zero crashes and one timeout ($+3.4\,\mathrm{pp}$, McNemar exact $p = 0.549$, $7$ vs.\ $4$ discordant pairs; Table~\ref{tab:tb21main}).
An initial sweep with always-on Turn-0 subagents enabled scored $72/89$ ($80.9\%$); the $79/89$ run reflects a second, tuned configuration with eager Turn-0 subagents disabled (chosen after RQ2) and the persistent \texttt{python\_repl} enabled (used on $19/89$ tasks for $82$ calls).
Under this setup, git-harness finishes faster than Pi on both median ($68.5\,\mathrm{s}$ vs.\ $72.9\,\mathrm{s}$) and mean ($194.6\,\mathrm{s}$ vs.\ $231.2\,\mathrm{s}$) duration.
It lowers mean API spend by $15.8\%$ ($\$0.238$ vs.\ $\$0.282$; total $\$21.16$ vs.\ $\$25.14$), while median per-task cost is slightly higher ($\$0.150$ vs.\ $\$0.141$).
Live KV-cache hit rates are statistically indistinguishable from Pi ($88.72\%$ vs.\ $87.39\%$ median, Wilcoxon $p = 0.525$; $94.08\%$ vs.\ $93.71\%$ weighted), consistent with treating append-only prefixes as a structural property of the harness, not a measured hit-rate advantage over Pi on this endpoint.
Git checkpointing adds only $0.55\%$ median wall time, and all $89/89$ trials export valid bundles ($42.5\,\mathrm{KB}$ median) even across timeouts.

\subsection{RQ2: Single-seed component ablation (15 feature-active tasks)}
\label{sec:rq2}
\begin{figure}[!t]
\centering
\includegraphics[width=\linewidth]{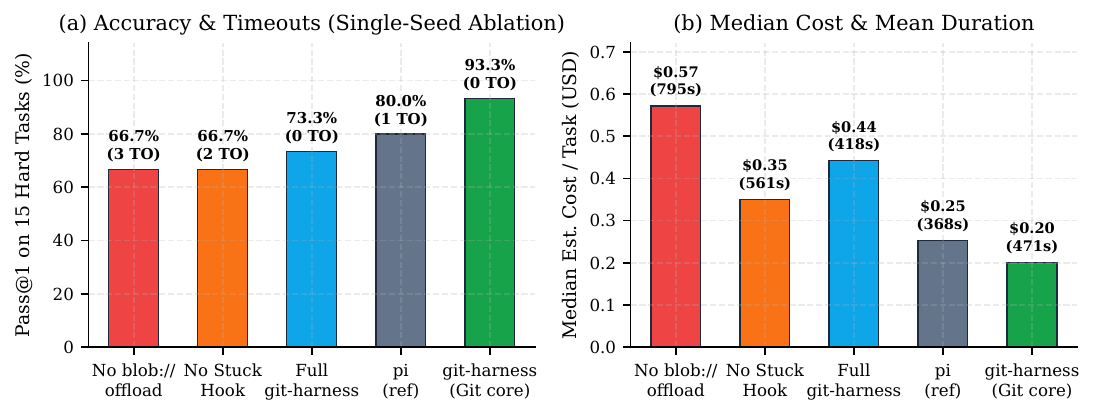}
\caption{\textbf{Single-seed 4-arm component ablation on 15 feature-active Terminal-Bench v2.1 tasks.} Removing \texttt{blob://} offloading or \texttt{StuckLoopHook} coincided with container timeouts and $+34\%\text{--}+90\%$ mean duration increases, whereas disabling eager Turn-0 \texttt{subagent} delegation lifted Pass@1 to $14/15$ ($93.3\%$) and cut median cost by $55\%$.}
\label{fig:ablation}
\end{figure}

To see how individual mechanisms behave on long tasks, we selected the 15 most feature-active tasks under $950\,\mathrm{s}$ from the initial full-harness 89-task sweep ($72/89$).
We then toggled off one mechanism at a time ($k=1$): blob offloading, stuck-loop guardrails, and always-on subagent delegation.
Because this ablation uses one seed ($n = 15$, $k = 1$) on tasks drawn from the same sweep, Pass@1 deltas are underpowered and timeout counts reflect single-run events (Table~\ref{tab:ablation}, Figure~\ref{fig:ablation}).

\begin{table}[!t]
\centering
\caption{Single-seed ($k=1$) 4-arm component ablation on the 15 most feature-active Terminal-Bench v2.1 tasks (plus \texttt{pi} reference). Costs include root + nested subagent tokens under 10:1 cache pricing.}
\label{tab:ablation}
\resizebox{\linewidth}{!}{%
\begin{tabular}{@{}lcccccc@{}}
\toprule
\textbf{Configuration} & \textbf{Pass@1} & \textbf{Timeouts} & \textbf{Med / Mean Dur} & \textbf{W-KV-Cache} & \textbf{Med / Mean Cost} & \textbf{Total Cost} \\
\midrule
\texttt{pi} (reference) & $12/15$ ($80.0\%$) & $1$ & $144.4\,\mathrm{s}$ / $367.8\,\mathrm{s}$ & $91.86\%$ & $\$0.253$ / $\$0.307$ & $\$4.60$ \\
Full \texttt{git-harness} & $11/15$ ($73.3\%$) & $0$ & $404.1\,\mathrm{s}$ / $417.7\,\mathrm{s}$ & $96.16\%$ & $\$0.442$ / $\$0.651$ & $\$9.76$ \\
\quad $-$ No \texttt{blob://} offload & $10/15$ ($66.7\%$) & \textbf{$3$} & $419.9\,\mathrm{s}$ / \textbf{$795.4\,\mathrm{s}$} & $96.69\%$ & \textbf{$\$0.571$} / \textbf{$\$0.904$} & \textbf{$\$13.56$} \\
\quad $-$ No \texttt{StuckLoopHook} & $10/15$ ($66.7\%$) & \textbf{$2$} & $429.7\,\mathrm{s}$ / \textbf{$561.3\,\mathrm{s}$} & $96.85\%$ & $\$0.350$ / $\$0.678$ & $\$10.17$ \\
\quad $-$ No \texttt{subagent} (Git core) & \textbf{$14/15$ ($93.3\%$)} & \textbf{$0$} & $296.5\,\mathrm{s}$ / $470.9\,\mathrm{s}$ & \textbf{$97.76\%$} & \textbf{$\$0.201$} / $\$0.649$ & $\$9.73$ \\
\bottomrule
\end{tabular}}
\end{table}

Disabling blob offloading nearly doubled mean wall time ($418\,\mathrm{s}$ to $795\,\mathrm{s}$), raised total spend by $39\%$, and coincided with three timeouts (Table~\ref{tab:ablation}).
This aligns with prior evidence for observation masking~\citep{lindenbauer2025complexity}.
Disabling stuck-loop guardrails raised mean duration by $34\%$ and coincided with two timeouts.
Disabling always-on subagents moved the opposite way.
Keeping only the Git core lifted Pass@1 from $11/15$ to $14/15$ with zero timeouts, raised weighted KV-cache hit rate to $97.8\%$, and cut median per-task cost by $55\%$ ($\$0.442$ to $\$0.201$).
This echoes Prime Agent's observation that current models struggle to decide when to spawn subagents~\citep{karten2026primeagent}.
We therefore ran git-harness with subagents disabled by default in RQ3.

\subsection{RQ3: Multi-model and cross-harness comparison}
\label{sec:rq3-matrix}

To test how harness designs interact across model families~\citep{zhang2026disclose,guo2026survey}, we evaluated five harnesses across four frontier models.
Four harnesses ran per model ($240$ matrix trials) on the same 15 Terminal-Bench v2.1 tasks in isolated Docker containers at high reasoning effort.
git-harness, Pi, and Prime Agent ran across all four models; Claude Code ran on Opus 5.5 and Fable 5.2; and Codex ran on self-hosted DeepSeek v4.1 and GLM 5.3.
We parsed all token and cost metrics directly from raw agent session logs.
Harbor's summary file undercounts Claude Code streaming output tokens by $48\times\text{--}62\times$ and zeroes usage fields on timeouts, omitting $\$19.97$ ($22.4\%$ of total Anthropic spend).
Because each cell covers $n = 15$ tasks ($1$ task $= 6.7\,\mathrm{pp}$), this matrix tests cross-model parity, cost behavior, and failure modes rather than fine-grained Pass@1 rankings.

\begin{table}[!t]
\centering
\caption{Four-model evaluation across five harnesses (four per model, $16$ cells $\times 15 = 240$ matrix trials, plus $31$ wire-protocol validation trials and $30$ latest-runtime Anthropic trials) on 15 hard Terminal-Bench v2.1 tasks. \textbf{TO}: container timeouts; \textbf{Med Dur}: median wall time. Anthropic dollar costs use effective unit rates inferred from \texttt{claude-code}'s cumulative \texttt{modelUsage} ledger (Opus 5.5: $\$4/\$20/\$0.20/\$5$; Fable 5.2: $\$8/\$40/\$0.20/\$10$ per 1M fresh-in/out/cache-read/cache-write; see Limitations).}
\label{tab:matrix}
\small
\begin{tabular}{@{}llcccr@{}}
\toprule
\textbf{Model (\texttt{high})} & \textbf{Harness} & \textbf{Pass@1} & \textbf{TO} & \textbf{Med Dur} & \textbf{Spend (\$/Pass)} \\
\midrule
\multirow{4}{*}{Claude Opus 5.5}
  & \texttt{git-harness}        & \textbf{$14/15$ ($93.3\%$)} & \textbf{$0$} & $280\,\mathrm{s}$ & $\$8.39$ ($\$0.60$) \\
  & \texttt{pi 1.7.0}           & $12/15$ ($80.0\%$)          & \textbf{$0$} & \textbf{$128\,\mathrm{s}$} & \textbf{$\$4.65$ ($\$0.39$)} \\
  & \texttt{prime-agent 0.10.5} & $13/15$ ($86.7\%$)          & $1^*$        & $146\,\mathrm{s}$ & $\$7.01$ ($\$0.54$) \\
  & \texttt{claude-code}        & $13/15$ ($86.7\%$)          & \textbf{$0$} & $264\,\mathrm{s}$ & $\$7.88$ ($\$0.61$) \\
\midrule
\multirow{4}{*}{Claude Fable 5.2}
  & \texttt{git-harness}        & $13/15$ ($86.7\%$)          & \textbf{$0$} & $849\,\mathrm{s}$ & $\$30.69$ ($\$2.36$) \\
  & \texttt{claude-code}        & \textbf{$14/15$ ($93.3\%$)} & $1$          & \textbf{$271\,\mathrm{s}$} & $\$22.54$ ($\$1.61$) \\
  & \texttt{prime-agent 0.10.5} & $12/15$ ($80.0\%$)          & \textbf{$0$} & $297\,\mathrm{s}$ & \textbf{$\$17.83$ ($\$1.49$)} \\
  & \texttt{pi 1.7.0}           & $12/15$ ($80.0\%$)          & $2^*$        & $382\,\mathrm{s}$ & $\$22.35$ ($\$1.86$) \\
\midrule
\multirow{4}{*}{DeepSeek v4.1}
  & \texttt{git-harness}        & \textbf{$12/15$ ($80.0\%$)} & $2^*$        & $824\,\mathrm{s}$ & local \\
  & \texttt{prime-agent 0.10.5} & $10/15$ ($66.7\%$)          & \textbf{$0$} & \textbf{$446\,\mathrm{s}$} & local \\
  & \texttt{pi 1.7.0}           & $9/15$ ($60.0\%$)           & $1$          & $590\,\mathrm{s}$ & local \\
  & \texttt{codex 0.77.0}       & $4/15$ ($26.7\%$)           & $4$          & $680\,\mathrm{s}$ & local \\
\midrule
\multirow{4}{*}{GLM 5.3}
  & \texttt{git-harness}        & $7/15$ ($46.7\%$)           & $2$          & \textbf{$249\,\mathrm{s}$} & local \\
  & \texttt{pi 1.7.0}           & $7/15$ ($46.7\%$)           & $2$          & $275\,\mathrm{s}$ & local \\
  & \texttt{prime-agent 0.10.5} & \textbf{$8/15$ ($53.3\%$)}  & $2$          & $322\,\mathrm{s}$ & local \\
  & \texttt{codex 0.77.0}       & $6/15$ ($40.0\%$)           & $5$          & $835\,\mathrm{s}$ & local \\
\bottomrule
\end{tabular}\par
\vspace{0.35em}
\raggedright\footnotesize $^*$Includes timeout trial(s) whose final workspace still passed pytest verification (both DeepSeek v4.1 \texttt{git-harness} timeouts passed pytest). On Claude Opus 5.5 and Claude Fable 5.2, \texttt{git-harness} ran with \texttt{--python-repl}, \texttt{--step-timeout-sec 300}, and zero timeouts; its sole Opus 5.5 failure is Anthropic's third-party Cyber Safeguards refusal on \code{feal-linear-cryptanalysis}.
\end{table}

\begin{figure}[!t]
\centering
\includegraphics[width=\linewidth]{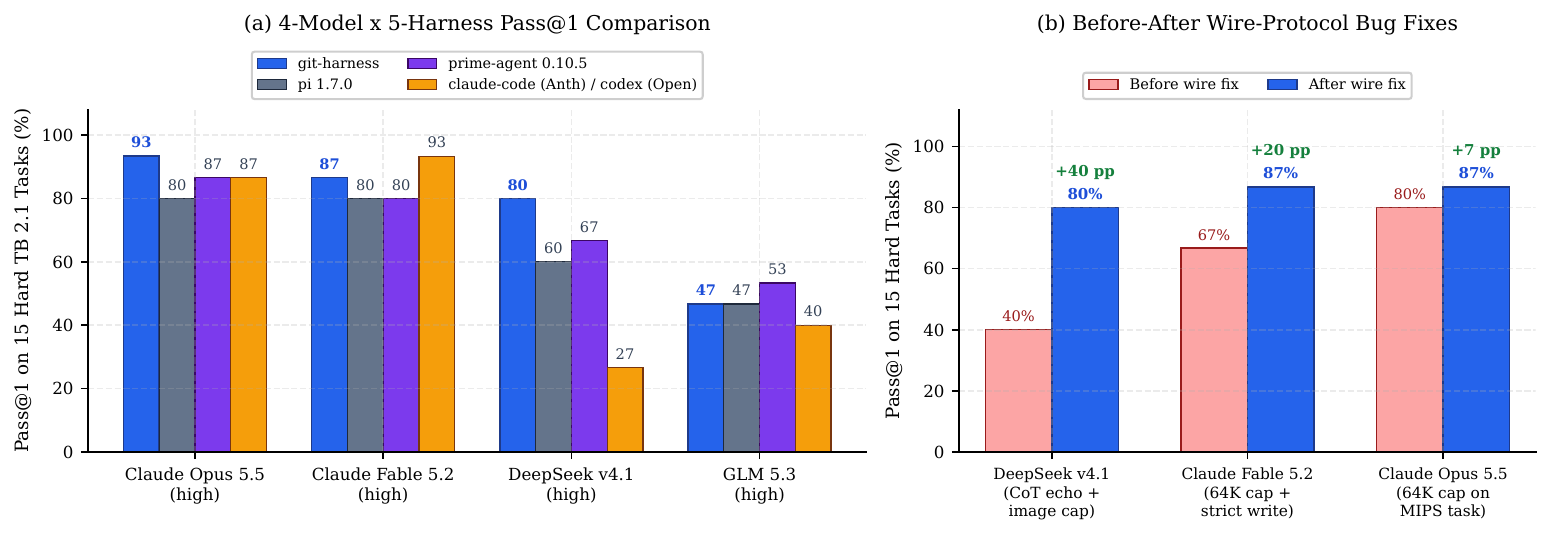}
\caption{\textbf{Four-model cross-harness Pass@1 and wire-protocol debugging deltas.} (a) Pass@1 across four frontier models and five harnesses (four per model, $n=15$ per cell); (b) before-after Pass@1 deltas from applying Lesson~4 wire-protocol fixes to \texttt{git-harness}.}
\label{fig:matrix-ab}
\end{figure}

\begin{figure}[!t]
\centering
\includegraphics[width=\linewidth]{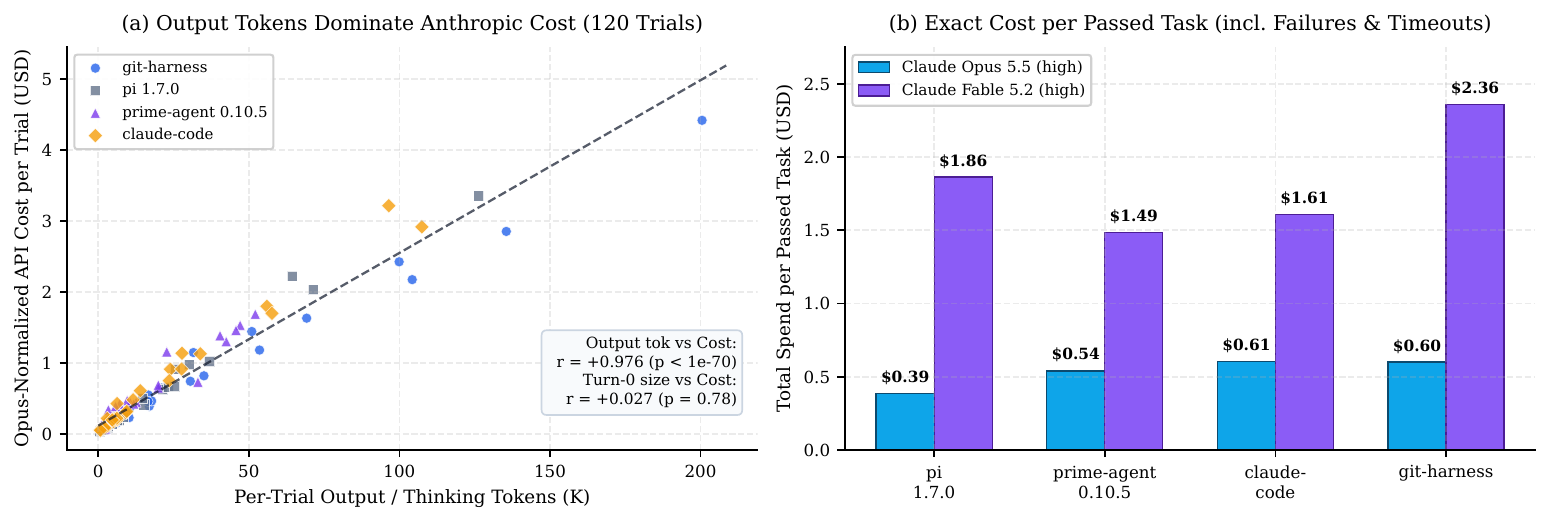}
\caption{\textbf{Anthropic prompt-cache and cost breakdown across 120 trials.} (a) Total spend scales almost linearly with output and thinking tokens ($r = +0.976$ on per-model price-normalized cost) because prompt-cache reads are $100\times\text{--}200\times$ cheaper than output tokens; (b) total spend per passed task including failures and timeouts.}
\label{fig:anth-cost}
\end{figure}

\paragraph{Anthropic prompt-cache behavior and cost per pass.}
Across 120 Anthropic trials (Table~\ref{tab:matrix}, Figures~\ref{fig:matrix-ab} and~\ref{fig:anth-cost}), raw session logs show all four harnesses achieving $89.0\%\text{--}93.2\%$ weighted prompt-cache hit rates.
Because cached input reads cost roughly two orders of magnitude less than output tokens, output and thinking volume drives almost all spend ($r = 0.976$ on price-normalized cost).
Initial prompt size has negligible effect ($r = 0.027$), even though Claude Code ($17.6\mathrm{K}\text{--}19.0\mathrm{K}$ median Turn-0 tokens) and Prime Agent ($13.3\mathrm{K}\text{--}13.4\mathrm{K}$) start with $6\times\text{--}9\times$ larger Turn-0 prompts than git-harness ($2.3\mathrm{K}$) and Pi ($2.8\mathrm{K}$).
Note that the Anthropic git-harness cells in Table~\ref{tab:matrix} reflect a re-run on our newer runtime (\texttt{--python-repl} and \texttt{--step-timeout-sec 300}), whereas the other harnesses kept their single initial matrix run.
In cost per pass, git-harness trades higher output volume for solving harder tasks.
On Opus 5.5, git-harness reaches $14/15$ ($93.3\%$) at $\$0.60$/pass and is the only harness to solve \code{filter-js-from-html}.
That cost per pass sits slightly below Claude Code ($\$0.61$) and above Prime Agent ($\$0.54$) and Pi ($\$0.39$).
On Fable 5.2, git-harness solves $13/15$ with zero timeouts, including \code{filter-js-from-html} for $\$2.36$.
Its average cost per pass on Fable 5.2 is $\$2.36$, the highest of the four harnesses (Prime Agent $\$1.49$, Claude Code $\$1.61$, Pi $\$1.86$), and its median duration ($849\,\mathrm{s}$) is roughly $3\times$ that of Claude Code ($271\,\mathrm{s}$) and Prime Agent ($297\,\mathrm{s}$).
Longer extended-thinking traces and more turns ($14.0$ median steps in raw logs vs.\ $7.0$ for Pi and $8.0$ for Claude Code) explain both the extra spend and wall time; \code{sam-cell-seg} alone used $200.5\mathrm{K}$ output tokens ($\$8.84$).

\paragraph{Tool usage patterns.}
Even with 12--22 registered tools in Claude Code, Anthropic models routed $81\%\text{--}91\%$ of calls through \texttt{Bash}.
They never called \texttt{Grep}, \texttt{Glob}, \texttt{Agent}, or \texttt{TodoWrite}.
Similarly, Prime Agent routed $98.7\%$ of calls through its persistent Python REPL and made zero recursive subagent (\texttt{rlm.spawn}) or online refinement (\texttt{refine.run}) calls across inspected trials.
Its strong results on GLM 5.3 ($8/15$) and DeepSeek v4.1 ($10/15$) come from stateful in-memory Python execution, not recursive subagent calls.

\paragraph{Provider API vs.\ first-party safeguards.}
On \code{feal-linear-cryptanalysis} ($n = 1$ task), all $6/6$ third-party API runs across Opus 5.5 and Fable 5.2 (git-harness, Pi, Prime Agent) terminated at Step~2.
Anthropic's Cyber Safeguards refused each run right after reading \texttt{feal.c}, whereas both first-party Claude Code runs proceeded and passed (Appendix~\ref{app:traces}, Case Study B).
Although observed on one task, this shows that provider safety policies can treat first-party and API clients differently on security-adjacent benchmarks.

\paragraph{Per-harness failure modes.}
Each harness showed recurring failure modes across models.
Codex suffered edit-and-test oscillation timeouts, including $127$ consecutive edits to one file on \code{make-mips-interpreter} and $113$ repeated test runs on \code{filter-js-from-html}.
Pi stalled inside blocking synchronous shell calls ($1{,}847\,\mathrm{s}$ on \code{regex-chess}).
Prime Agent deadlocked for $3{,}148\,\mathrm{s}$ on leaked async shell handles \emph{after} already passing the task's tests (Appendix~\ref{app:traces}, Case Study C).
git-harness exported valid bundles on all $60/60$ runs at $0.35\%$ mean Git overhead ($0.12\%$ excluding one nested \texttt{git clone} cache outlier).

\subsection{RQ4: Wire-protocol debugging lessons and unpacked DAG forensics}
\label{sec:rq4}

\paragraph{Before-after debugging deltas (Figure~\ref{fig:matrix-ab}b).}
Auditing portable bundles and per-step profiles across our four models turned up three wire-protocol bugs in earlier git-harness runs.
These are before-after bug fixes on our own harness, and on DeepSeek v4.1 two fixes shipped together.
They are practical debugging lessons, not single-factor controlled ablations.
First, dropping prior reasoning traces on DeepSeek v4.1 forced the model to re-derive state every turn and dropped Pass@1 to $6/15$ with $5$ timeouts.
Echoing prior reasoning and capping inline images at $4$ per compaction window lifted Pass@1 to $12/15$ and cut total output tokens by $14.4\%$.
On \code{video-processing}, bounding inline PNG frames also resolved a decode collapse caused by $30\text{--}57$ repeated images.
Second, capping Anthropic adaptive-thinking models at $16\mathrm{K}$ output tokens with forgiving \texttt{write} coercion caused truncated writes to fail silently.
This bug dropped Fable 5.2 from $13/15$ to $10/15$ and trapped Opus 5.5 on \code{make-mips-interpreter} in an $11$-turn empty-write loop ($180.2\mathrm{K}$ loop tokens; $223\mathrm{K}$ tokens and $\$4.81$ total) before timeout.
Raising the streaming ceiling and rejecting missing \texttt{write} content solved \code{make-mips-interpreter} in $551\,\mathrm{s}$ for $\$1.53$ (Appendix~\ref{app:traces}, Case Study A) and lifted Opus 5.5 from $12/15$ ($80.0\%$) to $13/15$ ($86.7\%$) in Figure~\ref{fig:matrix-ab}b. Adding \texttt{python\_repl} and the step watchdog later lifted Opus 5.5 to $14/15$ ($93.3\%$) in Table~\ref{tab:matrix}.

\paragraph{Unpacked DAG case studies.}
Appendix~\ref{app:traces} reproduces unpacked commit graphs from \texttt{session.bundle} alongside cross-harness side-by-side forensics.
In \textbf{\code{winning-avg-corewars}}, the parent spawns three concurrent subagents (\code{strategy-a/b/c}) that consume $1.18\mathrm{M}$ worker tokens across isolated branches.
Three dual-parent merge commits (\code{869c43da}, \code{64a967aa}, \code{42c612eb}) then integrate the results while keeping parent context under $13.5\mathrm{K}$ tokens.
In \textbf{\code{path-tracing-reverse}}, a $243$-step run ($247$ commits) maintains a $99.3\%$ live KV-cache hit rate.
At Step~224, a dual-parent compaction anchor (\texttt{c5a8756c}) drops active context from $160.0\mathrm{K}$ to $1.5\mathrm{K}$ tokens, followed by $2\text{--}4\,\mathrm{ms}$ FTS5 \texttt{recall} queries on Steps~239 and~241.

\section{Related Work}
\label{sec:related}

\paragraph{Coding agent harnesses and harness engineering.}
SWE-agent~\citep{yang2024sweagent} formalized Agent-Computer Interfaces (ACIs), while OpenHands~\citep{wang2024openhands,openhands2025sdk} introduced a modular event-stream runtime and SDK.
Recent empirical work treats harness engineering as its own discipline~\citep{barbaste2026harness,ning2026codeasharness,guo2026survey}.
Controlled factorials show that harness variance can exceed model variance by $7.8\times$~\citep{zhang2026disclose}, while longitudinal studies quantify how harness changes shift accuracy on fixed models~\citep{sghaier2026dontblame}.
Several recent systems also explore automated harness synthesis~\citep{lou2026autoharness,lee2026metaharness,lin2026agenticharness}.
Architecturally, \textbf{Prime Agent}~\citep{karten2026primeagent} is closest to our work.
It combines a persistent IPython REPL (RLM abstraction), async recursive subagents over live message queues, and a typed Continual Harness on custom disk persistence.
Prime Agent prioritizes programmatic expressivity and live messaging, while noting open issues with irreversible actions and contaminated online skill updates.
git-harness instead prioritizes a minimal, zero-dependency Git storage layer with atomic chat-and-file rewind, boundary-only branch merges, deterministic pre-commit gates, and append-only prompt prefixes (Table~\ref{tab:compare}).
We adopt Prime Agent's persistent Python REPL pattern as an optional Layer~2 tool while keeping all state, orchestration, and provenance strictly Git-native.

\begin{table}[!t]
\centering
\caption{Architectural comparison of git-harness against contemporary coding agent harnesses.}
\label{tab:compare}
\resizebox{\linewidth}{!}{%
\begin{tabular}{@{}lccccc@{}}
\toprule
\textbf{Capability} & \textbf{\texttt{git-harness}} & \textbf{Prime Agent} & \textbf{GCC} & \textbf{OpenHands} & \textbf{\texttt{pi}} \\
\midrule
Storage substrate & Native Git plumbing & Custom disk/daemon & \texttt{.GCC/} MD & Event log + FS & JSONL \\
Co-atomic chat + FS rewind & \textbf{Yes (1 commit)} & Separate & Partial & No & No \\
Zero host-repo pollution & \textbf{Yes (shadow idx)} & N/A & No (\texttt{.GCC/}) & Container & N/A \\
Append-only prompt prefix & \textbf{Yes (\texttt{-1st-par})} & Not reported & No & Event-stream & Append-only \\
Agent-to-agent comm. & \textbf{Branch merge + notes} & Async queues & Manual merge & Event bus & None \\
Multi-agent provenance & \textbf{\texttt{parent2} + notes} & Queue tree & Manual merge & Sub-events & None \\
Pre-commit memory gates & \textbf{Unit-tested} & Noted open issue & No & No & N/A \\
Portable single-file trace & \textbf{\texttt{git bundle}} & Log dir & Dir & Log dir & JSONL \\
\bottomrule
\end{tabular}}
\end{table}

\paragraph{Git-based agent state and memory.}
Git Context Controller (GCC)~\citep{wu2025gcc} models agent memory using Git metaphors (\texttt{COMMIT}, \texttt{BRANCH}, \texttt{MERGE}) inside a custom \texttt{.GCC/} markdown directory.
EvoGit~\citep{evogit2025} coordinates decentralized code evolution through a Git phylogenetic graph.
git-harness drops the metaphor and runs directly on Git's object database, private shadow indices, linked worktrees, and notes.
This unifies sessions, workspace snapshots, blob offloading, subagents, memory, and skills without touching user branches.

\paragraph{Context compression and KV-cache economics.}
ACE~\citep{zhang2026ace} evolves itemized context playbooks to prevent context collapse, while Lindenbauer et al.~\citep{lindenbauer2025complexity} show that observation masking matches LLM summarization at lower cost.
git-harness's blob offloading pairs observation masking with content-addressed retrieval.
First-parent projection keeps prompt prefixes append-only purely at the harness layer across any provider that supports prefix caching.

\paragraph{Evaluation rigor for coding agents.}
Terminal-Bench~\citep{merrill2026terminalbench} provides realistic containerized CLI tasks.
HAL~\citep{kapoor2026hal} and the Agentic Benchmark Checklist (ABC)~\citep{zhu2025abc} set standards for cost-normalized evaluation, paired significance testing, and transparent reporting, which we follow throughout Section~\ref{sec:evaluation}.

\section{Conclusion and Limitations}
\label{sec:conclusion}

git-harness offers an alternative way to structure a coding agent harness: putting all five subsystems directly on Git's content-addressed object model replaces five separate stores with one zero-dependency runtime.
Our evaluation tests whether this Git-native design holds up in practice alongside existing harnesses.
On the 89-task Terminal-Bench v2.1 sweep, git-harness matches Pi ($79/89$ vs.\ $76/89$, $p = 0.549$), lowers mean latency ($194.6\,\mathrm{s}$ vs.\ $231.2\,\mathrm{s}$) and mean cost ($\$0.238$ vs.\ $\$0.282$; median $\$0.150$ vs.\ $\$0.141$), adds $0.55\%$ median Git overhead, and achieves $100\%$ bundle survival.
Across 240 cross-harness trials spanning four frontier models and five harnesses, git-harness achieves comparable Pass@1 with zero Anthropic timeouts, while taking longer and costing more per pass on Fable 5.2 because of extended thinking traces.
Our trajectories also point to three practical lessons: eager Turn-0 subagent delegation hurt accuracy and cost in our single-seed ablation, output tokens and timeout tail-risk govern Anthropic spend, and portable forensics expose wire-protocol bugs that otherwise hide behind fixed-model scores.

\paragraph{Limitations.}
\label{sec:limitations}
Several boundaries scope these findings.
First, all Terminal-Bench v2.1 runs use a single seed ($k = 1$) and $n = 15$ cross-model slices, so small accuracy differences ($p = 0.549$ on the 89-task sweep) are underpowered. Both the $79/89$ sweep (up from $72/89$ with full subagents) and the Anthropic git-harness cells also reflect a second runtime pass on the same benchmark, not an untouched held-out suite.
Second, RQ1 used one OpenAI-compatible endpoint, Anthropic dollar figures rely on effective unit rates inferred from Claude Code's usage ledger (not a published rate card), and live KV-cache hits still depend on provider-side TTLs and eviction outside the harness.
Third, boundary-only subagent merges intentionally give up mid-flight peer steering (unlike queue-based systems such as Prime Agent~\citep{karten2026primeagent}), and Layer~4's memory promotion gates are covered by unit tests, not an adversarial memory-poisoning benchmark.
Addressing these bounds requires multi-seed evaluation of the frozen runtime on held-out benchmarks.

\bibliographystyle{plainnat}
\bibliography{references}

\appendix
\section{Unpacked \texttt{session.bundle} Forensic Traces \& Case Studies}
\label{app:traces}

Every Terminal-Bench v2.1 run exports a portable Git bundle (\texttt{session.bundle}) inspectable with standard Git commands.
Below are exact excerpts from real runs discussed in Section~\ref{sec:evaluation}.

\begin{figure}[!h]
\centering
\includegraphics[width=\linewidth]{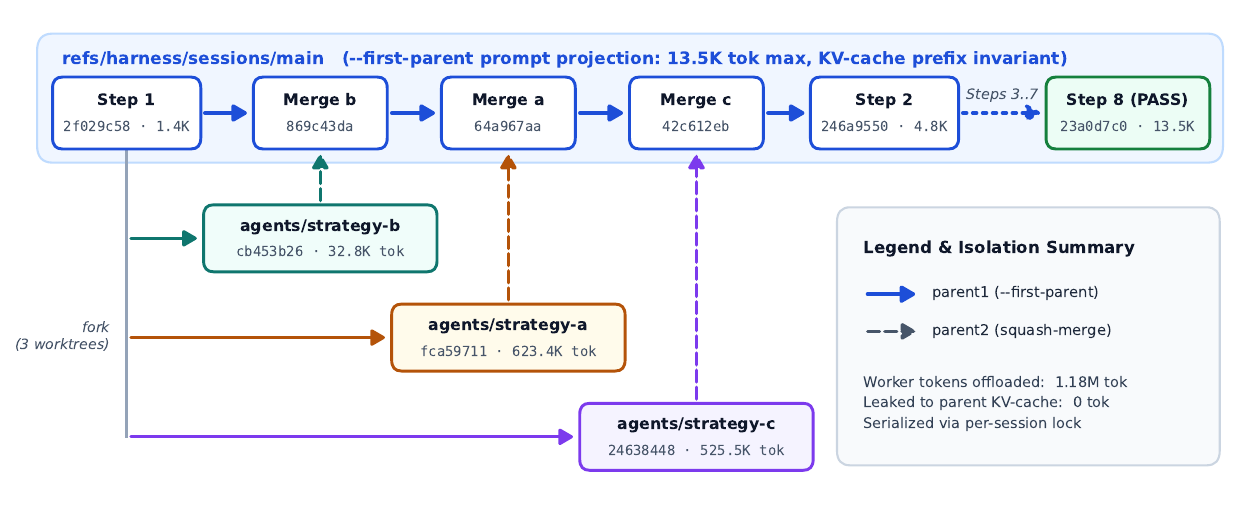}
\caption{\textbf{Unpacked Git DAG from \code{winning-avg-corewars} (\texttt{session.bundle}).} Three isolated worktree subagents fork from Step~1, consume $1.18\mathrm{M}$ worker tokens, and join the main session via serialized dual-parent squash-merge commits. Prompt projection follows only the solid \texttt{--first-parent} spine ($\le 13.5\mathrm{K}$ tokens), while dashed \texttt{parent2} edges preserve complete worker history in Git.}
\label{fig:dag-trace}
\end{figure}

\paragraph{A.1 Parallel Subagent Squash-Merge DAG (\code{winning-avg-corewars}).}
On Step~2, the parent spawned three concurrent worktree subagents (\texttt{strategy-a/b/c}) that consumed $1.18\mathrm{M}$ worker tokens across isolated branches.
Three serialized dual-parent \texttt{subagent-merge} commits then joined them onto \code{refs/harness/sessions/main} (Figure~\ref{fig:dag-trace}).
Walking \texttt{git rev-list --first-parent} follows only the solid top spine (\texttt{2f029c58} $\to$ \texttt{869c43da} $\to$ \texttt{64a967aa} $\to$ \texttt{42c612eb} $\to$ \texttt{246a9550} $\to \dots \to$ \texttt{23a0d7c0}).
This excludes all $1.18\mathrm{M}$ worker tokens from the parent KV-cache while preserving full provenance via \texttt{parent2}.

\begin{figure}[!h]
\centering
\includegraphics[width=\linewidth]{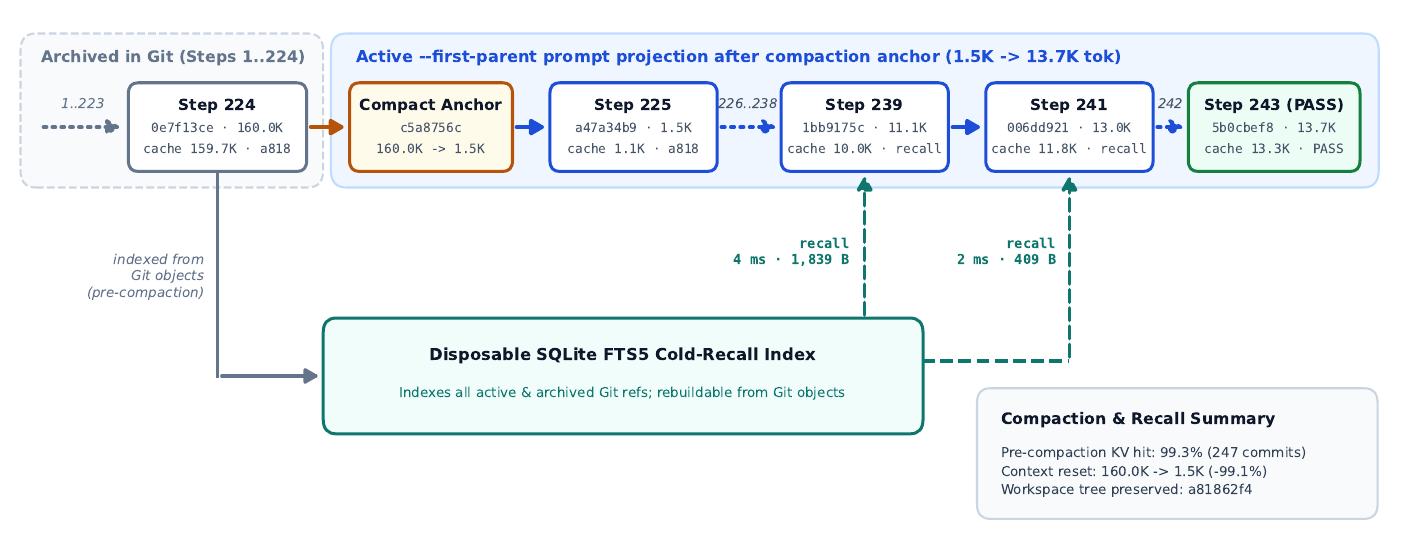}
\caption{\textbf{Unpacked compaction and cold-recall trace from \mbox{\texttt{path-tracing-reverse}} (\texttt{session.bundle}).} Compaction anchor \texttt{c5a8756c} resets active \texttt{--first-parent} projection from $160.0\mathrm{K}$ to $1.5\mathrm{K}$ tokens ($-99.1\%$) while leaving Steps~1--224 archived in Git. Subsequent steps recover archived details in $2\text{--}4\,\mathrm{ms}$ through the disposable SQLite FTS5 index.}
\label{fig:compact-trace}
\end{figure}

\paragraph{A.2 Dual-Parent Auto-Compaction + Cold Recall (\mbox{\texttt{path-tracing-reverse}}).}
At Step~224 (\texttt{0e7f13ce}), active prompt context reached $160.0\mathrm{K}$ tokens at a $99.3\%$ cumulative KV-cache hit rate.
Layer~1 committed compaction anchor \texttt{c5a8756c} and reset Step~225 (\texttt{a47a34b9}) to $1.5\mathrm{K}$ tokens while preserving workspace tree \texttt{a81862f4}.
Steps~239 (\texttt{1bb9175c}) and~241 (\texttt{006dd921}) then queried SQLite FTS5 \texttt{recall} in $4\,\mathrm{ms}$ and $2\,\mathrm{ms}$ to recover archived pre-compaction details (Figure~\ref{fig:compact-trace}).

\paragraph{A.3 Cross-Harness Forensic Case Studies.}
\begin{itemize}[leftmargin=*,nosep]
  \item \textbf{Case Study A: Output-cap truncation loop (Opus 5.5 on \code{make-mips-interpreter}).} The model attempted to write \texttt{/app/vm.js} in one $\sim 15.5\mathrm{K}$-token step. With a $16\mathrm{K}$ output cap and forgiving \texttt{write} coercion, the tool call truncated $11$ times in a row and silently wrote $0$ bytes with no error signal. That loop burned $180.2\mathrm{K}$ output tokens ($223\mathrm{K}$ tokens and $\$4.81$ across the full trial) before timeout. Raising the streaming output cap and validating \texttt{write} arguments solved the task in $551\,\mathrm{s}$ for $\$1.53$.
  \item \textbf{Case Study B: Provider API vs.\ first-party safeguards (\code{feal-linear-cryptanalysis}, $n=1$ task).} On Opus 5.5 and Fable 5.2, first-party Claude Code solved the task in $4\text{--}5$ steps. In contrast, $6/6$ third-party API runs (git-harness, Pi, Prime Agent) ended at Step~2 under Anthropic's Cyber Safeguards refusal right after reading \texttt{feal.c}. Recording explicit refusal stop reasons prevents policy blocks from masquerading as silent completions.
  \item \textbf{Case Study C: Async shell handling (Opus 5.5 on \code{regex-chess}).} All four harnesses achieved \texttt{reward=1}, yet wall time spanned $6.8\times$ ($526\,\mathrm{s}$ to $3{,}600\,\mathrm{s}$). git-harness finished fastest ($526\,\mathrm{s}$), and Claude Code ($872\,\mathrm{s}$) explicitly cleaned up background fuzzers via \texttt{pkill}. Pi ($1{,}994\,\mathrm{s}$) spent $93\%$ of its time stalled inside blocking synchronous shell calls. Prime Agent passed \texttt{check.py} in $451\,\mathrm{s}$ but then deadlocked for $3{,}148\,\mathrm{s}$ on six un-killed async shell handles until timeout.
\end{itemize}

\section{LLM Usage Disclosure}
Large language models assisted with implementation, benchmark automation, and manuscript formatting. All experimental protocols, statistical analyses, architectural claims, and empirical measurements were executed and verified directly against real test suites and Terminal-Bench v2.1 runs.

\end{document}